# The CTTC 5G end-to-end experimental platform
## Integrating heterogeneous wireless/optical networks, distributed cloud, and IoT devices

Raul Muñoz, Josep Mangues, Ricard Vilalta, Christos Verikoukis, Jesus Alonso-Zarate, Nikolaos Bartzoudis, Apostolos Georgiadis, Miquel Payaró, Ana Perez-Neira, Ramon Casellas, Ricardo Martínez, José Núñez-Martínez, Manuel Requena-Esteso, David Pubill, Oriol Font-Bach, Pol Henarejos, Jordi Serra, Francisco Vazquez-Gallego, CTTC

*The IoT will facilitate a wide variety of applications in different domains such as smart cities, smart grids, industrial automation (Industry 4.0), smart driving, elderly assistance, or home automation, among others. Billions of heterogeneous smart devices with different application requirements will be connected to the networks and will generate huge aggregated volumes of data that will be processed in distributed cloud infrastructures. On the other hand, there is also a general trend to deploy functions as software instances in cloud infrastructures (e.g., NFV or MEC). Thus, the next generation of mobile networks, the so-called 5G, will need not to only develop new radio interfaces or waveforms to cope with the expected traffic growth but also to integrate heterogeneous networks from end-to-end with distributed cloud resources to deliver end-to-end IoT and mobile services. This paper presents the first-known end-to-end 5G platform that is being developed by CTTC, capable of reproducing such ambitious scenario.*

## Introduction

The fifth generation of mobile networks technology (5G) is not only about the development of new radio interfaces or waveforms. It also deals with the design of end-to-end converged network and cloud infrastructure to facilitate both traditional human-based and emerging Internet of Things (IoT) services. This converged infrastructure, illustrated in Fig. 1, is composed of: *i)* end-to-end heterogeneous network segments covering radio and fixed access, metro aggregation, and core transport with heterogeneous wireless and optical technologies; *ii)* massive distributed cloud computing and storage infrastructures; and *iii)* large amounts of heterogeneous smart devices and terminals for traditional mobile broadband services (e.g., smartphones, tablets, etc.) and IoT services (e.g. sensors, actuators, robots, cars, drones, etc.).

At the network level, the requirements for 5G include high flexibility, low-latency, and high-capacity in order to support the forecasted 1000x growth in mobile data traffic with sub-millisecond latency [1]. On the control/management side, 5G networks also require to deliver end-to-end (E2E) connectivity services between distributed cloud infrastructures and between any end user (i.e., devices and terminals) and the distributed cloud infrastructure. This requirement can only be met by efficiently integrating heterogeneous access (RAN, fixed access, satellite, Wi-Fi, personal area networks), optical/wireless crosshaul (fronthaul/backhaul), metro aggregation packet networks and high-capacity optical core transport networks.

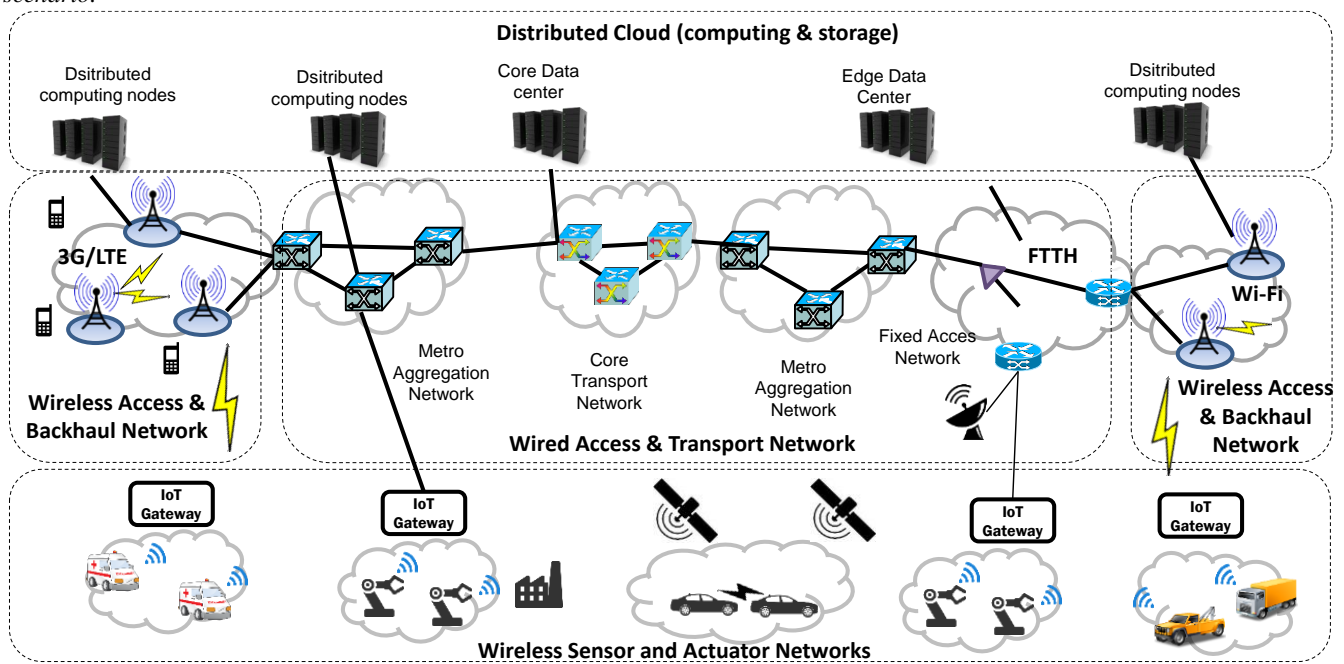

**Figure 1** 5G end-to-end scenario integrating distributed cloud, heterogeneous networks, IoT and mobile broadband services.

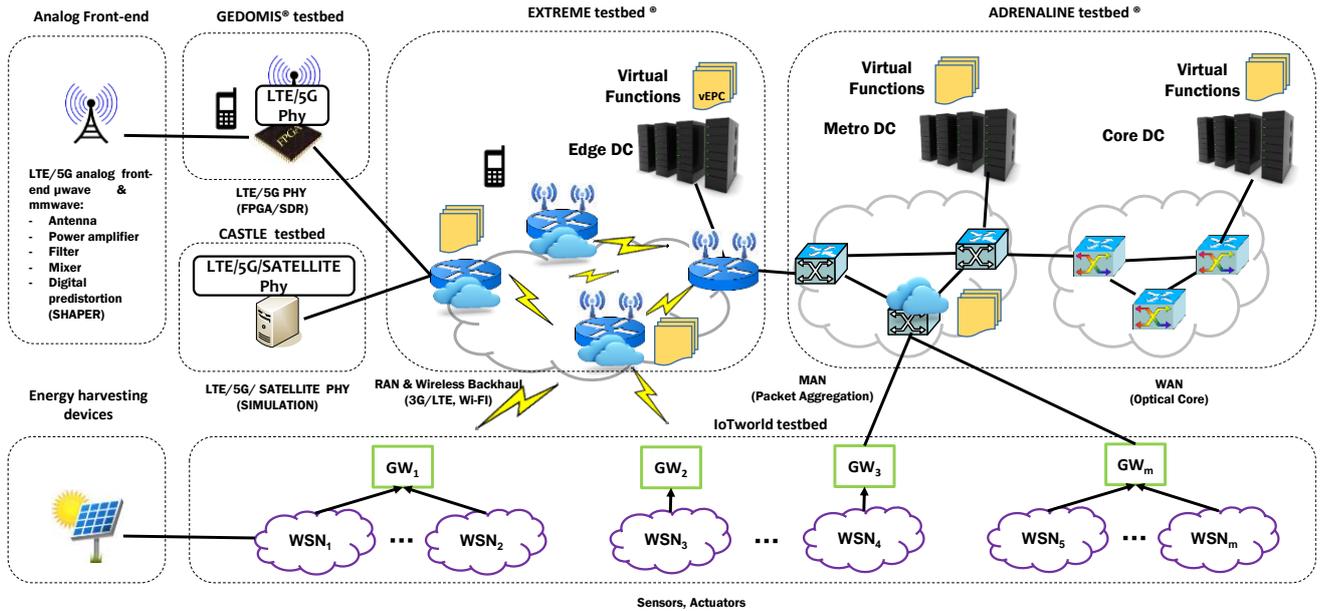

*Figure 2 CTTC 5G end-to-end experimental platform.*

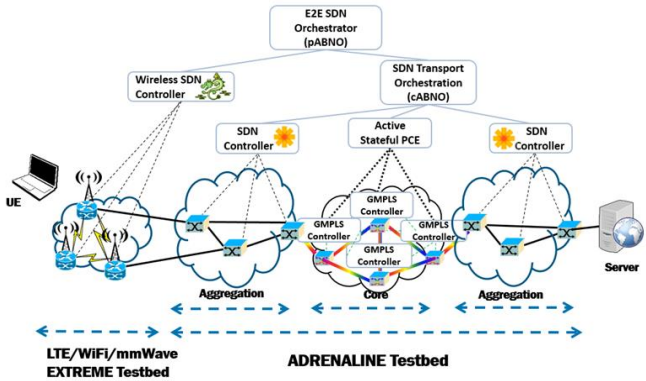

*Figure 3 Integration of ADRENALINE and EXTREME testbeds.*

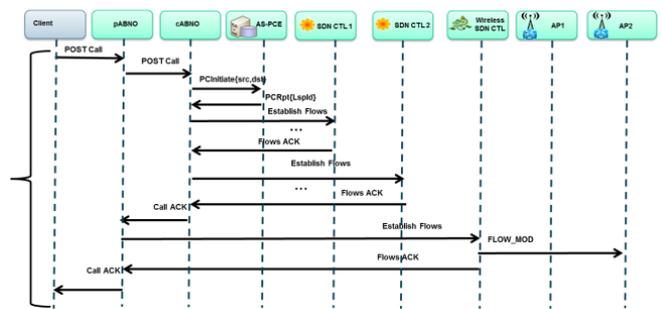

*Figure 4 Message exchange diagram of E2E provisioning services between ADRENALINE and EXTREME Testbeds.*

At the cloud level, 5G requires massive computing and storage infrastructures. This is needed to store and process (e.g. Big Data) the data being generated by billions of smart devices connected collecting information like temperature monitoring, distance measurement, energy consumption, etc. In addition, the impending growth of Network Function Virtualization (NFV) [2] and Mobile Edge Computing (MEC) [3] also require cloud services for the deployment of software functions (e.g., mobile Evolved Packet core – EPC-, local cache, firewalls, etc.,). Originally, cloud services have been implemented in core data centers (DCs) for high-computational or long-term processing. However, the cloud is being spread to the edge of the network (e.g., in edge DCs located in the metro network, or even in network nodes or mobile base stations with cloud capabilities) in order to reduce the latency of services for the end user. This concept is referred to as *fog computing*. Therefore, 5G networks need a global orchestration for the distributed cloud implementation and the management of heterogeneous networks. This orchestration shall dynamically allocate computing and storage resources to deploy functions where needed, and provide the required connectivity in order to achieve the desired end-to-end functionality or service.

Conducting real-life demonstrations of such a complex system is not easy. CTTC is working in the development of the first-known 5G end-to-end experimental platform for testing advanced end-to-end IoT and mobile services. The approach consists in integrating various existing experimental facilities already available at CTTC which cover activities from the PHY layer to the application/service layer for mobile networks. The building blocks of this demonstration platform are shown in

Fig.2. These facilities cover complementary technologies ranging from terminals, sensors and machines, to radio access networks, aggregation/core networks, and cloud/fog computing. Specifically, the five existing experimental facilities involved are namely: *i)* the ADRENALINE Testbed® [4] for wired fronthaul/backhaul (SDN-enabled packet aggregation and optical core network, distributed cloud and NFV services in core and metro data-centers); *ii)* the EXTREME Testbed® [5] and LENA LTE-EPC protocol stack emulator for wireless fronthaul/backhaul and mobile core (SDN-enabled wireless HetNet and backhaul, edge data center and distributed computing nodes for cloud and NFV services); *iii)* the GEDOMIS® testbed [6] for LTE/5G PHY real-time prototyping based on FPGAs and software-defined radio (SDR), and the CASTLE testbed (a highly configurable software tool allowing LTE/5G/Satellite PHY layer development and testing); *iv)* an LTE/5G analog front-end μwave & mmwave (antenna, power amplifier, filter, mixer, etc.,) including digital pre-distortion (SHAPER), and energy harvesting devices for the IoT; and *v)* the IoTWorld Testbed [7] integrating sensors, actuators, and wireless/wired gateways. The aim of this paper is to describe how all these facilitates are integrated with each other to build a complete end-to-end (E2E) 5G demonstration infrastructure.

## Integrating wireless and optical transport networks for E2E connectivity

Hierarchical SDN Orchestration has been proposed as a feasible solution to handle the heterogeneity of different network domains, technologies, and vendors. It focuses on network control and abstraction through several control domains, whilst using standard protocols and modules. The need of hierarchical SDN orchestration has been previously justified with two purposes: scaling and security.

Fig. 3 shows the proposed hierarchical SDN architecture for the integration of wireless and optical transport networks [8]. In the wireless segment, implemented over the EXTREME Testbed®, an SDN controller is in charge of the programming of the wireless network (access and backhaul). This SDN controller tackles the specificities of the wireless medium, implementing the proper extensions to control wireless devices. In the optical segment, implemented over the ADRENALINE Testbed®, we consider an SDN-enabled MPLS-TP aggregation network, while a core network uses an Active Stateful PCE (AS-PCE) on top of a GMPLS-controlled optical network.

In the proposed architecture, we introduce a Global E2E SDN Orchestrator, which is responsible for the provisioning of E2E connections through different network segments. It has been implemented using the parent ABNO (pABNO) in [9]. The pABNO is able to orchestrate several network segments: an SDN-enabled wireless segment and an optical transport network segment controlled by a child ABNO (cABNO). ABNO is an IETF RFC which describes the internals of an SDN Controller.

Fig. 4 shows the proposed message exchange between a pABNO and a wireless SDN controller/cABNO. It can be observed that an E2E connection is requested (POST Call) to the pABNO. The pABNO computes the involved network controllers (Wireless SDN/cABNO) and requests the underlying connection to them. We can observe how the workflow follows inside a cABNO, which is responsible for another level of hierarchical SDN orchestration.

## Integrating the IoT and Optical metro Aggregation Networks with Distributed Cloud computing

SDN is a key enabler technology to address all the technical challenges posed by the IoT. SDN aims to overcome the limitations of traditional IP networks, which are complex and hard to manage in terms of network configuration and reconfiguration due to faults and changes. SDN can be viewed as a network operating system which interacts with the data plane and the network applications by means of Application Programmable Interfaces (APIs). In this regard, also the different needs in networking resources such as bandwidth and delay can be managed more easily thanks to the software programmability approach facilitated by SDN in the network control. Another important benefit of SDN is that it paves the way for the integration of smart objects with fog and cloud computing. More specifically, thanks to the flexibility provided by SDN, the data flows of information between IoT nodes and fog or cloud computing can be easily managed. This enables collaborative analytics between geo-distributed smart things.

Integrating IoT and SDN can also increase the efficiency of the network by responding to changes or events detected by the IoT which might imply network reconfiguration. For example, SDN can be used in IoT applications where the data are transmitted from the sensors periodically in specific time frames to schedule the requested bandwidth on the transmission paths only during the active duty cycles. Such dynamic reconfiguration of the forwarding devices is only possible via centralized applications which orchestrate IoT collected information and network resources information jointly. SDN security can also be applied to IoT gateways in order to enforce security at the network edges.

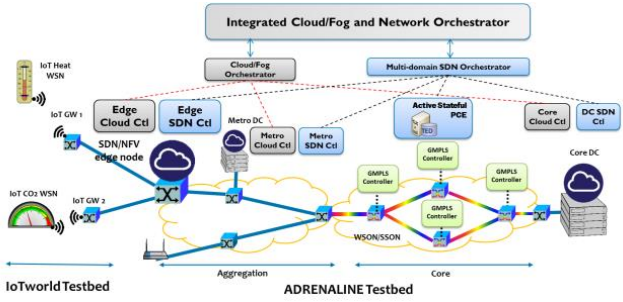

*Figure 5. End-to-End SDN Orchestration of SDN/NFV-enabled Edge Node for IoT Services*

We have deployed an SDN/NFV-enabled Edge Node in ADRENALINE Testbed® for integrating wired IoT gateways from the IoTWorld Testbed by means of E2E SDN Orchestration of integrated Cloud/Fog and network resources [10]. E2E SDN orchestration will provide network connectivity between IoT gateways and deployed virtual machines (VMs) which might be allocated in the proposed edge node or in a DC located in the core network.

Fig. 5 shows the considered system architecture. At the top of the figure, the Integrated Cloud/Fog and Network Orchestrator [11] is responsible for handling Virtual Machine (VM) and network connectivity requests, which are processed through the Cloud and SDN orchestrators. The orchestration process consists of two different steps: the VM creation and network connectivity provisioning. The integrated Cloud/Fog and Network Orchestrator requests the creation of virtual instances (VMs) to the Cloud Orchestrator, which, is responsible for the creation of the instances. It is also responsible to attach the VMs to the virtual switch inside the host node (at the edge node or in a core DC). When the VMs creation is finished, the Cloud Orchestrator replies the VM's networking details to the integrated Cloud/Fog and network orchestrator (MAC address, IP address and physical computing node location). The SDN orchestrator is the responsible to provision E2E network services. The SDN orchestrator will provide the E2E connectivity between the requested IoT gateway and the deployed VM. Finally, data from IoT gateway will flow to the processing resources located in the proposed SDN/NFV-enabled edge node.

Fig. 6 shows the cloud and network topology as seen by the Cloud/Fog and Network Orchestrator. Each network domain (network elements controlled by a single SDN controller) are abstracted as a single node (either packet or optical). It can also be observed that the provisioned VMs are connected to the packet network.

## Integrating the IoT with wireless access and backhaul networks

Experimenting with cellular networks, including the wireless backhaul, and the IoT is a challenging venture. A powerful alternative for research purposes would consist in using Software Define Radios (SDR) to deploy an IoT network which is LTE-enabled. This would mean that every single IoT device is equipped with an SDR LTE radio and connects with an SDR-based eNodeB. Then the various eNodeBs could be interconnected using another SDR-based solution. Unfortunately, both the technical complexity and the cost of designing, deploying, and operating such a complex solution

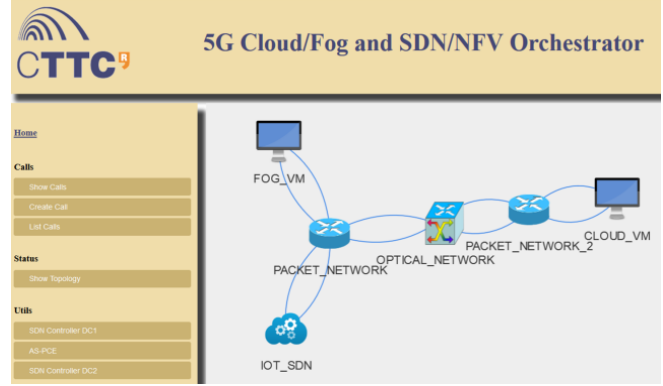

*Figure 6. View of the cloud and network resources from the developed cloud/fog and network orchestrator*

would become prohibitive, especially if the number of devices of the IoT network has to be large.

As an alternative, the solution presented in the E2E Testbed of CTTC consists in connecting real IoT devices to EXTREME Testbed® in various ways. Towards this end, it is first necessary to integrate the category-0 UE in the LENA modules. Such integration would then enable the following levels of connection between the IoTWorld and EXTREME Testbed®:

1) First option would consist in running LENA in the gateways, e.g. Raspberry-Pi modules of IoTWorld. LENA could operate in emulation mode with an instance of a UE fetching the data packets actually received via radio, routing them through the protocol stack of LENA, and emulating their transmission to the eNodeB. The EPC would also be emulated in LENA, thus offering an E2E solution. This approach is shown in Figure 7.

2) The main limitation of the first option is that it would only be possible to run a single gateway connected to a single eNodeB, thus limiting the capability of the joint testbed to emulate a realistic IoT network where more than one Machine-to-Machine (M2M) Gateway is expected to be deployed at the same time to run an IoT application. In order to overcome this limitation, the second option would consist in setting up a central server where LENA runs with various instances of UEs. Then, the data traffic actually received by each gateway via radio, i.e. raspberry-pi, would be connected to each of the UE instances of the server. Such interconnection would enable the emulation of various M2M gateways simultaneously connected to a common eNodeB and the EPC. This approach is shown in Figure 8. With any of the above two approaches, the IoTWorld testbed could also benefit from the existing connections between EXTREME and ADRENALINE, thus enabling for a complete E2E experimentation of an IoT deployment. In this case, data would be transported via SDN-controlled wireless and optical backhaul networks acting as heterogeneous multi-domain transport layer, as shown in Figure 8.

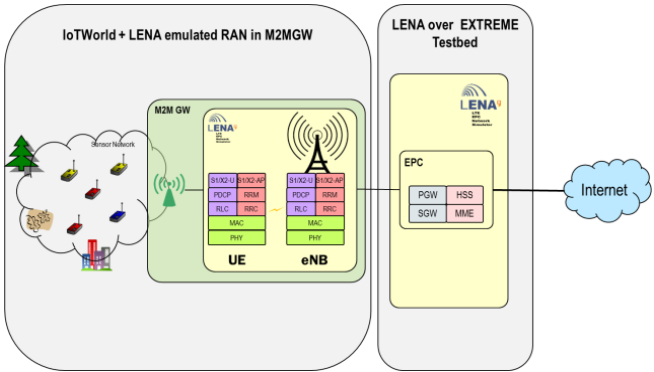

**Figure 7** *M2M Gateway connected to UE Stack of LENA running inside a single LENA process that includes the eNodeB and the Core Network.*

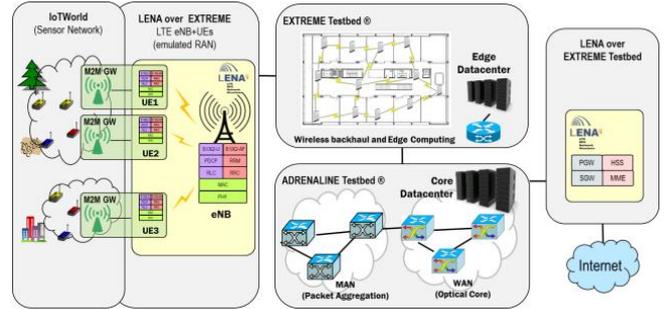

**Figure 8.** *End-to-End MTC Testbed with emulated RAN and EPC with wireless and optical backhaul.*

## Integrating 5G/LTE Phy-layer and virtual mobile protocol stack for flexible HW/SW partitioning

The integration of a 5G/LTE PHY-layer provided by either the GEDOMIS testbed or CASTLE testbed with the LENA LTE-EPC emulated protocol stack [12] running over the EXTREME Testbed will allow full-stack experimentation of multiple 5G use cases that exploit the flexibility of SDN/NFV when applied to mobile networks (e.g., virtual base stations or self-organized networking). This may consider, for example, the virtual base station use case defined by the ETSI NFV group, or experiments on coverage and capacity optimization involving the whole radio protocol stack and its interactions with core network elements and real applications. The availability of full-stack mobile network testbed allowing experimentation from PHY up to applications and services is currently rare. In fact, experiments on the above topics are often limited to either 1) PHY layer platforms with minimal MAC layer support, due to the cost of commercial protocol stacks, and the limitations of open source ones, or 2) IP-level testbeds with limited access to low-level PHY configuration. In contrast, our integrated testbed will allow evaluating full-stack NFV solutions in a real wireless propagation environment with GEDOMIS testbed, or in a real-time emulation environment with CASTLE emulator, with the possibility of combining both emulated and real (e.g., fiber) backhaul/aggregation network links and applications, and of including additional emulated cells in order to achieve a larger experiment scale.

Fundamental to realize this vision is to develop the tools allowing to exploit the flexibility that NFV brings to mobile networks. In this sense, it is important to underline that the flexible HW-SW partitioning is widely perceived nowadays as a key technology enabler for fully exploiting the network programmability brought by SDN and NFV. In this way, the software programmable data plane combined with intelligent hardware that dynamically collaborates with control plane SW, will allow addressing the performance, flexibility, and security challenges of 5G mobile networking. At the same time, the requirements for low latency communications (including signal processing), low power operation and high performance parallel computation will be made more stringent, thus imposing dedicated hardware solutions for some functions. There are three main issues related to the HW-SW split i) the physical location where communication stacks are implemented (e.g., base station, cloud), ii) the technology of the processing elements that is used (e.g., entirely programmable system-onchip -SoC-, FPGAs, general purpose processors -GPP-) and iii) the granularity of the partitioning (e.g., at communication stack level, at function or process level). In this use case, the main idea is: 1) to flexibly distribute functions of the communication stack among different processing elements (e.g., SoCs, FPGAs, GPPs), which are physically residing either in communication nodes (e.g., base stations) or in cloud/fog processing architectures, and 2) to adaptively change processing topologies (e.g., base stations, intermediate nodes or cloud processing architectures). The decision for the distribution of the HW-SW functions will be made based on experiments made at design time (static mode), which will reflect specific 5G operating scenarios, as shown in Fig. 9.

Such function distribution may be done based on multiple criteria, such as high performance, low energy, low communication latency or a tradeoff of the previous. Apart from shifting functions from dedicated hardware co-processors to software processing space to guarantee the performance goal of the identified scenarios, a number of baseband parameters will be tunable. The tuning of these parameters will be performed at system level and, potentially, could also be based on different criteria. Initially, for this particular use case, the decisions will be driven by energy aware communications criteria. Essentially, these system level decisions will be based on using different communication primitives that can adjust certain parameters associated with a given energy cost. Indicative parameters of this type might be: i) the modulation and coding scheme, ii) the use of single or multi-antenna configurations (e.g., improve SNR or data rate), iii) the signal bandwidth, iv) the type of waveform (e.g., from 4G to 5G ones), v) the use of fragmented spectrum (e.g., coexistence of radio transmissions) [13], vi) the type of digital predistortion and crest factor reduction applied for the linearization of the power amplifier output (e.g., affecting the baseband processing requirements, DAC and ADC usage and power amplifier energy cost), vii) the energy consumption operation mode (e.g., switch off dynamic power consumption in portions of the FPGA baseband implementation), and viii) the traffic volume density allowing to switch off parts of the network. Fig. 10 shows two example of virtual mobile network function splitting and deployment developed in GEDOMIS and EXTREME testbed.

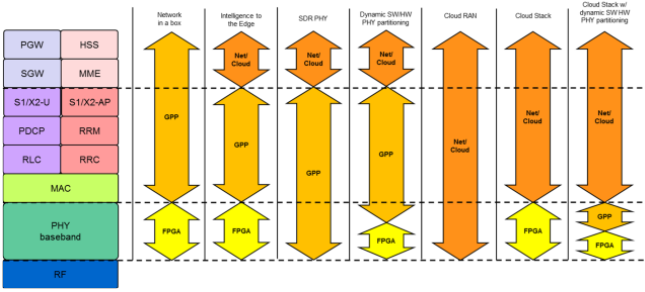

*Figure 9.* Virtual mobile network function splitting and deployment

## Integrating analog front-end and energy harvesting devices

Wireless communication is ultimately demonstrated through analog front-ends suitable to address the stringent requirements and diverse features of the physical layer digital signal processing of 5G systems acquired through GEDOMIS and CASTLE test-beds. At the same time, energy harvesting technologies exploiting existing light, thermal, vibration and electromagnetic ambient energy availability can further extend the energy autonomy of low power, Internet-of-Things communication, sensor and actuator devices, and improve the energy efficiency of the underlying analog front-ends and digital circuitry [14][15].

CTTC's test-bed facilities span a diverse set of fabrication technologies including traditional printed circuit board milling but also laser prototyping and inkjet printing able to provide the necessary resolution for low cost prototyping of analog front-ends up to millimeter wave frequencies and using different materials from flexible substrates suitable for wearables and Internet-of-Things wireless sensor devices, to high performance substrate materials suitable for millimeter wave applications. In addition, testing capabilities include signal generators, vector network analyzers, and an anechoic chamber suitable for testing wireless front-ends under controlled propagation environments. The use of off-the-shelf circuit components, and traditional microwave substrate materials and fabrication methods allow one to obtain low cost easily customized front-end designs, and additionally they permit a fast turnaround time of high performance prototypes.
In particular, the development and testing of analog front-ends capable to interface the digital baseband output from

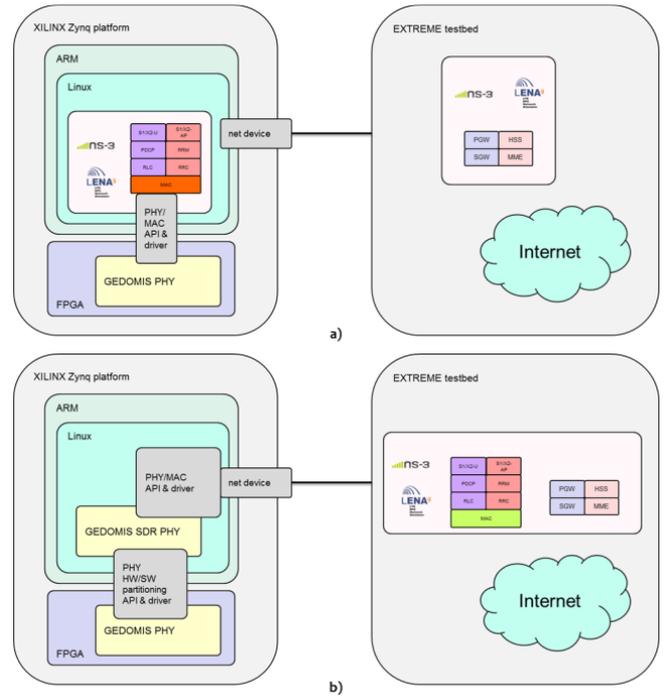

*Figure 10* a) Intelligence to the Edge, b) Cloud Stack with dynamic HW/SW PHY partitioning

GEDOMIS and CASTLE testbeds and introduce the necessary operating frequency translation and efficient amplification and filtering of the underlying signals into reconfigurable antenna arrays with beam-forming capabilities is of particular interest as it ultimately permits the real world testing of 5G systems and, more importantly, it enables the joint optimization of the baseband circuitry and analog front-end, which is not always possible when utilizing commercially available solutions. Similarly, the ability to implement and integrate different energy harvesters with the analog front-end, enables one to explore different circuit architectures and materials, low cost and mechanically conformal designs which can be jointly optimized and tailored to efficiently power wireless sensor circuits.

## Conclusions

Conducting real-life demonstrations of an end-to-end 5G scenario including both IoT and mobile broadband services, requiring the integration of heterogeneous wireless access and optical transport networks, distributed cloud computing, and wireless sensor and actuators networks is a very challenging task. CTTC has been working on the development of the first-known end-to-end 5G platform capable of reproducing such an ambitious scenario. This paper has described the existing and planned integration, supported functionalities, use cases, and preliminary results among the different experimental facilities available at CTTC.

**References**

[1] 5GPPP white paper, the 5G Infrastructure Public Private Partnership: the next generation of communication networks and services, March 2015, https://5g-ppp.eu/wp-content/uploads/2015/02/5G-Vision-Brochure-v1.pdf.

[2] Network function virtualization (nfv): Architectural framework," ETSI GS NFV 002 v.1.1.1,, 2013. http://www.etsi.org/deliver/etsi_gs/nfv/001_099/002/01.01.01_60/gs_nfv002v010101p.pdf

[3] Mobile-Edge Computing – Introductory Technical White Paper, ETSI MEC ISG, September 2014. https://portal.etsi.org/Portals/0/TBpages/MEC/Docs/Mobile-edge_Computing_-_Introductory_Technical_White_Paper_V1%2018-09-14.pdf

[4] "ADRENALINE Testbed," http://networks.cttc.es/ons/adrenaline/

[5] "EXTREME Testbed," http://networks.cttc.es/mobilenetworks/extreme_testbed/

[6] "GEDOMIS Testbed," http://technologies.cttc.es/phycom/gedomis/

[7] IoTWorld Testbed, http://iotworld.cttc.es.

[8] R. Vilalta et al., Hierarchical SDN Orchestration of Wireless and Optical Networks with E2E Provisioning and Recovery for Future 5G Networks. Optical Fiber Conference (OFC 2016), Anaheim, CA, USA, 2016.

[9] R. Vilalta *et al.*, Hierarchical SDN Orchestration for Multi-technology Multi-domain Networks with Hierarchical ABNO, European Conference on Optical Communication (ECOC 2015), Valencia, Spain, September 2015.

[10] R. Vilalta et al., End-to-End SDN Orchestration of IoT Services Using an SDN/NFV-enabled Edge Node, Optical Fiber Conference (OFC 2016), Anaheim, CA, USA, 2016.

[11] A. Mayoral, *et al.*, Experimental Seamless Virtual Machine Migration Using an Integrated SDN IT and Network Orchestrator, Optical Fiber Conference (OFC 2015), Los Angeles, CA, USA, 2015.

[12] N. Baldo, M. Miozzo, M. Requena, and J. N. Guerrero, "An open source product-oriented LTE network simulator based on ns-3," in In Proc. Of ACM MSWiM, 2011.

[13] O. Font-Bach, N. Bartzoudis, X. Mestre, D. López, P. Mege, L. Martinod, V. Ringset, T. André Myrvoll, "When SDR meets a 5G candidate waveform: providing efficient use of fragmented spectrum and interference protection for PMR networks", IEEE Wireless Communications, Vol. 22, No. 6, Dec. 2015.

[14] Sangkil Kim; Vyas, R.; Bito, J.; Niotaki, K.; Collado, A.; Georgiadis, A.; Tentzeris, M.M., "Ambient RF Energy-Harvesting Technologies for Self-Sustainable Standalone Wireless Sensor Platforms," in Proceedings of the IEEE , vol.102, no.11, pp.1649-1666, Nov. 2014

[15] Niotaki, K.; Collado, A.; Georgiadis, A.; Sangkil Kim; Tentzeris, M.M., "Solar/Electromagnetic Energy Harvesting and Wireless Power Transmission," in Proceedings of the IEEE, vol.102, no.11, pp.1712-1722, Nov. 2014.

**Acknowledgements**

This paper was partially supported by the European Commission through the FP7 EU-JP project STRAUSS (Contract Number 608528), H2020-ICT-2014-2 project Flex5Gware (Contract Number 671563) and 5G-Crosshaul (Contract Number 671598), the Spanish Ministry of Economy and Competitiveness (MINECO) through the projects FARO (TEC2012-38119), AEThER (TEC2014-58341-C4-4-R), CellFive (TEC2014-60130-P), 5GNORM (TEC2014-60491-R), and SOSRAD (TEC2012-39143), and by the Generalitat de Catalunya under grants 2014 SGR 1551.

**Raul Muñoz** (raul.munoz@cttc.es) *(SM'12) is graduated in telecommunications engineering in 2001 and received a Ph.D. degree in telecommunications in 2005, both from Universitat Politècnica de Catalunya (UPC), Spain. He is Head of the Optical Networks and Systems Department and Senior Researcher. Since 2000, he has participated in 36 R&D projects funded by the European Commission's Framework Programmes (H2020, FP7, FP6, FP5, CELTIC and ITEA), the Spanish research programmes, and the industry. Currently, he is the European project coordinator of the coordinated EU-Japan FP7 project STRAUSS (www.ict-strauss.eu). He has published over 45 journal papers, 160 international conference papers, and 1 patent.*

*Josep Mangues-Bafalluy received the MSc (1996) and PhD (2003) in Telecommunications from the Technical University of Catalonia (UPC). He is senior researcher and head of the communication networks division (networks.cttc.cat) of the Centre Tecnològic de Telecomunicacions de Catalunya (CTTC) and coordinated its IP Technologies Area from 2003 to 2012. He was also assistant professor at UPC. He regularly participates in EU (FP6, FP7, H2020), industrial (e.g., Cisco, Orange), Spanish, and Catalan research projects, of which six of them were led by him. His research interests include software-defined networks (SDN) and Network Functions Virtualization (NFV) applied to self-organized mobile networks.*

*Ricard Vilalta (ricard.vilalta@cttc.es) (M'12) graduated in telecommunications engineering in 2007 and received a Ph.D. degree in telecommunications in 2013, both from the Universitat Politècnica de Catalunya (UPC), Spain. Since 2010, Ricard Vilalta is a researcher at CTTC, in the Optical Networks and Systems Department. He is a Research Associate at Open Networking Foundation. He has been involved in international, EU and national research and industry projects. He has authored more than 90 journals and conference papers.*

*Christos Verikoukis*
http://www.cttc.es/people/cverikoukis/

*Jesus Alonso-Zarate (www.jesusalonsozarate.com), MSc. And PhD in Telecommunications Engineering, is Head of the Machine-to-Machine Communications Department at CTTC. Since 2010, he has published more than 130 technical publications in the area of wireless communications for the Internet of Things, focusing on the topics of Medium Access Control (MAC), scheduling, and energy efficiency. In the last years, he has received various best paper awards in international conferences and peer-reviewed journals. He is member of various editorial boards and, since April 2015, he is Editor-in-Chief of the Transactions on the Internet of Things endorsed by the European Alliance for Innovation.*

**Nikolaos Bartzoudis** is a senior researcher and head of the PHYCOM department at CTTC. He received his Ph.D. degree in electronic engineering from Loughborough University (UK 2006). Before joining CTTC (January 2008), Nikolaos has worked as Senior Research Officer (University of Essex, 2005-2007). He is currently involved (technical leader) with baseband development and experimental validation of spectral and energy efficient 4G-5G wireless communication technologies. His research interests include high-speed digital design, hardware-software co-design and reconfigurable SDR. In the past he was the principal investigator in 2 R&D projects and participated in 10 others (industrial, FP7/H2020, EPSRC). Currently he coordinates the AEThER project.

*Apostolos Georgiadis* (S'94–M'03-SM'08) received the Ph.D. degree in electrical engineering from the University of Massachusetts, Amherst, in 2002. In 2002 he joined GCD, USA as a Systems Engineer working on WLAN CMOS transceivers. In Mar. 2007 he joined CTTC, Spain as a senior researcher and since Apr. 2013 he is leading the Department of Microwave Systems and Nanotechnology. He is Associate Editor of the IEEE Microwave and Wireless Components Letters. He is Editor-in-Chief of the Wireless Power Transfer Journal. He is a Marie Curie Fellow and a Distinguished Lecturer of IEEE CRFID. He is Vice-Chair of URSI Commission D.

*Miquel Payaró* (SM'13) received the Ph.D. degree from the Universitat Politècnica de Catalunya in 2007. From February 2007 to December 2008 he was a Research Associate at the Hong Kong University of Science and Technology. Since January 2009 he is with CTTC where, in 2013, he became a Senior Researcher and was appointed the Head of the Communications Technologies Division. Miquel has led several research contracts with the industry and also has participated in publicly funded research projects (FP7, H2020), e.g., BuNGee, BeFEMTO or Newcom#. From July 2015, Miquel is serving as Technical Manager in the 5G-PPP project Flex5Gware (www.flex5gware.eu).

*Ana Pérez-Neira* is full professor at UPC (Technical University of Catalonia) in the Signal Theory and Communication department. She has been in the board of directors of ETSETB (Telecom Barcelona) and Vicerector for Research. Currently, she is Scientific Coordinator at CTTC (Centre Tecnològic de Telecomunicacions de Catalunya). Since 2008 she is member of EURASIP BoD (European Signal Processing Association) and since 2010 of IEEE SPTM (Signal Processing Theory and Methods). She has been the leader of 20 projects and has participated in over 50. She is author of 50 journal papers and more than 200 conference papers (20 invited).

*Ramon Casellas* (SM'12) graduated in telecommunications engineering in 1999 by both the UPC- university and ENST Paris, within an Erasmus double degree program. After working as an undergraduate researcher at France Telecom R&D and British Telecom Labs, he completed a Ph.D. degree in 2002. He worked as an associate professor at the ENST and joined the CTTC in 2006, with a Torres Quevedo grant. He currently is a senior research associate, involved in several international R&D and technology transfer projects. His research interests include network control and management, the GMPLS/PCE architecture and protocols, software defined networking and traffic engineering.

*Ricardo Martínez* (SM'14) graduated and PhD in telecommunications engineering by the UPC-BarcelonaTech university in 2002 and 2007, respectively. He has been actively involved in several public-funded (national and EU) R&D as well as industrial technology transfer projects. Since 2013, he is Senior Researcher of the Communication Networks Division (CND) at CTTC. His research interests include control and network management architectures, protocols and traffic engineering mechanisms for next-generation packet and optical transport networks within aggregation/metro and core segments.

*José Núñez-Martínez* is a Researcher in the Mobile Networks Department at CTTC in Barcelona. He received the Computer Science degree in 2004 and the PhD in 2014 at the Computer Architecture Depart. of the Technical University of Catalonia (UPC). From 2004 till 2007, he worked as a Network Engineer in the Advanced Broadband Communication Center (CCABA). He joined CTTC in 2007 as software network engineer. He has participated in several national (Cicyt), European (FP7), and industrial contracts (Ditech and AVIAT). He is author and co-author of more than 30 research papers.

**Manuel Requena Esteso** is M.Sc. in Computer Science from the Technical University of Valencia (UPV, 1998). In 1998, he realized his master thesis in the ENSEA (France). He worked as Software Engineer developing Telecommunication Solutions (AtosOrigin, France, 1998 - 2003). Currently, he is Research Engineer and coordinator of the EXTREME Testbed in the Mobile Networks Department of the CTTC. Since 2003, he has been participating in industrial and public-funded projects in wireless, mobile and optical networking: LENA LTE simulator (Ubiquisys/Cisco), VideoBond (DITEC), FREEDOM (Orange), BeFEMTO (ICT-FP7), WIP and NOBEL (IST-FP6), ARTICO and SYMBIOSIS (Spanish research projects) and TBONES (EUREKA).

David Pubill
http://www.cttc.es/people/dpubill/

***Oriol Font-Bach*** received his M.Sc. degree in Computer Engineering from Universitat Autònoma de Barcelona (UAB) in 2004 and the Ph.D. at the Department of Signal Theory and Communications of the Universitat Politècnica de Catalunya (UPC) in 2013. In 2006 he also obtained a Master in Design of Intregrated Circuits from UAB. He joined the CTTC in November 2006, where he currently holds a Researcher position in the PHYCOM department. He has participated in several national and European R&D projects. His research interests include FPGA-assisted SDR, hardware-software co-design and dynamically reconfigurable PHY-layer implementations towards energy and spectral efficient 5G systems was born in Barcelona. He received Telecommunication Engineering degree in May 2009. He joined the CTTC in January 2010 in Engineering area. In 2012 he obtained the Research on Information Technologies Master degree. He worked prototyping the physical layer communication technologies using software development. He also has expertise in LTE-A technologies thanks to industrial contracts for implementing the physical layer of LTE-A standard. Currently he is working on new technologies based on the implementation of the IEEE Standard 802.15.7 visible light communications in real devices and the Inmarsat BGAN.

*Jordi Serra*
<http://www.cttc.es/people/jserra/>

***Francisco Vázquez-Gallego*** received his B.Sc. in Electronics Engineering (1995) and his M.Sc. degree in Telecommunication Engineering (1998) from the Universitat Politècnica de Catalunya (UPC). He has more than 10 years of experience in analog and digital electronics R&D and project management, working in multidisciplinary projects at companies like NTE S.A., Tinytronic, Nub3D, Endesa and Amper. He participated in the development and manufacturing of prototypes and flight equipment for the European Space Agency (ESA). He has a broad and proven experience in the design and implementation of System-on-Chip on programmable devices. Since January 2010, he is working at the CTTC.